\documentclass[pre,aps,twocolumn,superscriptaddress,floatfix]{revtex4}

\usepackage{mathtools}
\usepackage[usenames,dvipsnames]{xcolor}
\usepackage{amsmath}
\usepackage{amssymb,mathrsfs}
\usepackage{graphicx}
\usepackage[colorlinks=true]{hyperref}
\usepackage{soul}
\usepackage{xcolor}

\usepackage[utf8]{inputenc}
\usepackage[english]{babel}

\newcommand{\al}{\alpha}

\newcommand{\la}{\lambda}
\newcommand{\ga}{\gamma}

\newcommand{\prt}{\partial}

\begin{document}

\title{Dynamics of Interaction of Two Soliton Clouds\footnote{JETP, {\bf 135,} 768 (2022)}
}

\author{A. M. Kamchatnov}
\affiliation{Institute of Spectroscopy,
Russian Academy of Sciences, Troitsk, Moscow, 108840, Russia}
\author{D. V. Shaykin}
  \affiliation{Moscow Institute of
  Physics and Technology, Institutsky lane 9, Dolgoprudny, Moscow
  region, 141700, Russia}

\begin{abstract}
On the basis of relationship between the kinetic equation for two soliton clouds in the theory 
of the Korteweg-de Vries equation and equations of the Chaplygin gas dynamics it is shown 
that the existence of waves propagating without a change in their form is a fundamental property of the 
nonlinear dynamics of soliton gases. The solutions of several typical problems in the soliton gas dynamics are 
considered and characteristic features of such dynamics, which make it possible to estimate the effects of 
interaction of soliton gases, are indicated.
\end{abstract}



\maketitle

\section{Introduction}

It is well known that the term ``solitons'' has been
introduced \cite{zk-65} analogously to the names of elementary
particles (electron, proton, etc.) in view of their elastic
interaction with one another for an important class of
nonlinear wave equations. Namely, two solitary waves
spaced by a large distance prior to their ``collision'',
return to their initial form after their passage through
the interaction stage without forming any additional
waves. Nevertheless, the event of their interaction is
not quite traceless: as a result of this interaction, the
trajectories of solitons acquire additional shifts as
compared to their initial trajectories \cite{lax-68}. For example,
in the case of the Korteweg-de Vries (KdV) equation,
which is a universal wave equation for describing
waves with account for weak nonlinearity and
dispersion effects and which can be written in terms of
standard dimensionless variables in form
\begin{equation}\label{eq1}
  u_t+6uu_x+u_{xxx}=0,
\end{equation}
the soliton solution is given by the expression
\begin{equation}\label{t6-23.3}
  u(x,t)=\frac{\kappa^2}2\cdot\frac1{\cosh^2[\kappa(x-\kappa^2t-x_0)/2]},
\end{equation}
i.e., velocity $s=\kappa^2$ of a soliton is proportional to its
amplitude $a=\kappa^2/2$.
If the wave excitation at the initial
instant can be represented to a high degree of accuracy
in the form of two soliton pulses(\ref{t6-23.3}) separated by a
large distance (the faster soliton with parameter $\kappa=\kappa_2$ has initial coordinate
$x_{02}$ on the left of coordinate $x_{01}$ of the slower soliton with parameter $\kappa_1$
($\kappa_1<\kappa_2$), their initial trajectories $x=\kappa_1^2t+x_{01}$,
$x=\kappa_2^2t+x_{02}$ acquire after the collision the shifts $x=\kappa_1^2t+x_{01}+\Delta x_1$,
$x=\kappa_2^2t+x_{02}+\Delta x_2$, where
\begin{equation}\label{t6-33.49}
  \Delta x_1=-\frac2{\kappa_1}\ln\frac{\kappa_2+\kappa_1}{\kappa_2-\kappa_1},\quad
  \Delta x_2=\frac2{\kappa_2}\ln\frac{\kappa_2+\kappa_1}{\kappa_2-\kappa_1}.
\end{equation}
Therefore, the fast soliton with a large amplitude is
shifted in the forward direction, while the slow and
lower soliton is shifted in the backward direction, the
shift of the lower soliton being larger (in absolute
value) than the shift of the higher soliton. Remarkably,
for an important class of integrable equations, solutions can be obtained with any number of 
solitons. In this case, as a result of simultaneous collision of three
or more solitons, the total shift of each soliton is equal
to the sum of shifts of form (\ref{t6-33.49})
for pairwise collisions (see, for example, \cite{zmnp-80,newell}), i.e., 
multiple simultaneous collisions of solitons in some spatial region do not differ in
this sense from sequential spatially separated pair collisions of solitons with one another.

Let us now suppose that in our wave system, a very
large number of solitons, each is characterized by its
parameter $\kappa$, are excited in the system. In this case, we
can speak of a soliton gas and use the concepts of gas
kinetics for its description. If we denote by $f(\kappa, x,t)dx d\kappa$ 
the number of solitons with coordinates from
interval $(x,x+dx)$, which have parameter $\kappa$ from interval 
$(\kappa,\kappa+d\kappa)$ at instant $t$, the evolution of such a
gas is characterized by the time dependence of distribution function $f(\kappa,x,t)$ 
in the coordinate and parameter $\kappa$. The problem of obtaining the kinetic equation
for such evolution was formulated by Zakharov in \cite{zakh-71} and was solved by him for a 
rarefied gas of solitons when their mutual collisions change velocities $s=\kappa^2$ very little. 
Later, El, who investigated the special limit of Whitham modulation equations for 
the infinite-phase solution of the KdV equation, obtained in \cite{el-03} a generalization 
of the Zakharov kinetic equation to a dense soliton gas. A simpler derivation of this 
equation based on the self-consistent determination of the mean velocity of a soliton moving
through the gas was given in \cite{elkamch-05} where the simplest
solution to these equations, which described the collision of two soliton gas clouds with close 
parameters $\kappa$ was also obtained. The formal solutions for any number of such clouds were 
analyzed in \cite{ekpz-11,fp-22}. At present,
the theory of kinetic equations for solitons has become
a part of actively developing ``generalized hydrodynamics'', 
which is applicable to any integrable models
of systems of many interacting particles (see, for example, reviews \cite{doyon-20,el-21} 
and references therein). However, despite such a remarkable formal progress,
the physical explanation of the behavior of soliton
gases appears as insufficient since the number of
solved problems that give an idea of characteristic features of the dynamics of interacting 
soliton gases is very scarce. At the same time, experimental investigations of soliton gas 
dynamics in various physical systems including waves in water and cold atoms have
been launched (see, for example,  \cite{redor-19,suret-20,bd-22}); this
requires the development of well-elaborated theory of
the soliton gas dynamics. In this article, we give several
examples of such typical dynamics, which clarify some
essential features of the soliton gas dynamics, distinguishing it from the dynamics of conventional gases.

\section{Kinetic equation for a soliton gas}

Let us first consider for completeness the brief derivation of the KdV kinetic equation for a soliton gas
following the method proposed in \cite{elkamch-05}. It should be
noted above all that in view of the integrability of the
KdV equation discovered in \cite{ggkm-67}, the evolution of wave $u(x,t)$
occurs so that the spectrum of the eigenvalue
problem for the time-independent Schrödinger equation
$\psi_{xx}=-(u(x,t)+\la)\psi$, which is associated with the
KdV equation, is independent of time $t$ and each soliton corresponds to a certain 
value $\la<0$ of discrete spectrum. More convenient parameter $\kappa$,
which has been used in the soliton solution in form (\ref{t6-23.3}), is connected with
$\la$ by relation $\kappa=\sqrt{-\la}$. Consequently,
during the evolution of the wave, both the spectrum
and the set of $\kappa$-values for the motion of solitons
accompanied with their collisions remain unchanged
in accordance with the KdV equation. 

If solitons did
not experience shifts (\ref{t6-33.49}) during their collisions, their
velocities could be expressed by formula $s=\kappa^2$
for all solitons with parameter $\kappa$. However, collisions modify
this velocity. In each collision of a soliton with slower solitons characterized by parameter
$\eta<\kappa$, the ``probe'' $\kappa$-soliton advances in the forward direction through
additional distance $\frac2{\kappa}\ln\frac{\kappa+\eta}{\kappa-\eta}$, and the number of such collisions per unit time is
equal to relative velocity $s(\kappa)-s(\eta)$, multiplied by the density of
$\eta$-solitons. Consequently, owing to
such collisions, the $\kappa$-soliton acquires additional velocity
$$
\int_0^{\kappa}\frac2{\kappa}\ln\frac{\kappa+\eta}{\kappa-\eta}\cdot
[s(\kappa)-s(\eta)]f(\eta)d\eta.
$$
Analogously, each soliton with parameter $\eta>\kappa$, which overtakes the probe
$\kappa$-soliton, shifts it per unit time backwards by distance
\begin{equation}
\begin{split}
&-\int_{\kappa}^{\infty}\frac2{\kappa}\ln\frac{\eta+\kappa}{\eta-\kappa}\cdot
[s(\eta)-s(\kappa))]f(\eta)d\eta=\\
&=\int_{\kappa}^{\infty}\frac2{\kappa}\ln\left|\frac{\kappa+\eta}{\kappa-\eta}\right|\cdot
[s(\kappa)-s(\eta)]f(\eta)d\eta,
\end{split}\nonumber
\end{equation}
which can be obtained by multiplying shift $-\frac2{\kappa}\ln\frac{\eta+\kappa}{\eta-\kappa}$ 
during a single collision by the number of such collisions $[s(\eta)-s(\kappa))]f(\eta)$ 
and integrating with respect
to values of $\eta>\kappa$. Adding these corrections to the
unmodified value of velocity $\kappa^2$, we obtain the following self-consistent equation:
\begin{equation}\label{t6-34.1}
  s(\kappa)=\kappa^2+\frac2{\kappa}\int_0^{\infty}
  \ln\left|\frac{\kappa+\eta}{\kappa-\eta}\right|[s(\kappa)-s(\eta)]f(\eta)d\eta.
\end{equation}
Therefore, the distribution function for solitons
with parameter $\kappa$ is transferred along the $x$ axis with
velocity $s(\kappa)$, defined by integral equation (\ref{t6-34.1}),
and the condition of the spectrum conservation during the
evolution of the wave in accordance with the KdV
equation can be written in the form of conservation law
\begin{equation}\label{t6-35.2}
  \frac{\prt f(\kappa,x,t)}{\prt t}+\frac{\prt[s(\kappa)f(\kappa,x,t)]}{\prt x}=0.
\end{equation}
This equation, supplemented with integral equation (\ref{t6-34.1}), 
is known as the KdV kinetic equation for a soliton gas; this equation was obtained by El in \cite{el-03}
using a different method. It should be noted that the inclusion in Eq.~(\ref{t6-34.1}) 
of pair collisions alone for a dense soliton gas is justified by the aforementioned addition of
shifts for multiple soliton collisions.

If the soliton gas is rarefied, i.e., $\int f(\kappa)d\kappa\ll\kappa_0$, where
$\kappa_0$ is the characteristic value of parameter $\kappa$ in
distribution $f(\kappa)$, the correction term in formula (\ref{t6-34.1}) is
small, and we can substitute unmodified value of $s(\kappa)\approx\kappa^2$ into it:
\begin{equation}\label{t6-35.3}
   s(\kappa)=\kappa^2+\frac2{\kappa}\int_0^{\infty}
  \ln\left|\frac{\kappa+\eta}{\kappa-\eta}\right|(\kappa^2-\eta^2)f(\eta)d\eta.
\end{equation}
This expression defines the function $s(\kappa)$ by a
closed expression in contrast to integral equation (\ref{t6-34.1}) for a dense soliton gas. 
System (\ref{t6-35.2}), (\ref{t6-35.3})
was obtained by Zakharov in \cite{zakh-71} together with the approach to the
kinetics of the soliton gas formulated here. Clearly,
such kinetic equations can also be obtained for other
completely integrable equations with the isospectral
evolution of nonlinear waves.

\section{Two-stream flows of soliton gases}

To get an idea of the dynamics of soliton gases based on kinetic equation
(\ref{t6-34.1}), (\ref{t6-35.2}), we assume that the
distribution function has two very narrow peaks near $\kappa_1$ and $\kappa_2$, 
i.e., this function can be written in form
\begin{equation}\label{t6-35.4}
  f(\kappa,x,t)=f_1(x,t)\delta(\kappa-\kappa_1)+f_2(x,t)\delta(\kappa-\kappa_2).
\end{equation}
This means that we consider the dynamics of interacting gases over times, 
such that we can disregard the
collisions of solitons of the same species with very
close velocities and take into account only collisions
between solitons of different species. Then the substitution of distribution function 
(\ref{t6-35.4}) into Eqs.~(\ref{t6-35.2}) and (\ref{t6-35.3})
gives conservation laws
\begin{equation}\label{t6-35.5}
  \frac{\prt f_1}{\prt t}+\frac{\prt (s_1f_1)}{\prt x}=0,\quad
  \frac{\prt f_2}{\prt t}+\frac{\prt (s_2f_2)}{\prt x}=0,
\end{equation}
where velocities $s_1,s_2$ satisfy equations
\begin{equation}\label{t6-36.6}
  s_1=\kappa_1^2+\alpha_1f_2(s_1-s_2),\quad s_2=\kappa_2^2+\alpha_2f_1(s_2-s_1),
\end{equation}
and we have introduced the following notation for convenience:
\begin{equation}\label{t6-36.7}
  \alpha_1=\frac2{\kappa_1}\ln\left|\frac{\kappa_1+\kappa_2}{\kappa_1-\kappa_2}\right|,\quad
  \alpha_2=\frac2{\kappa_2}\ln\left|\frac{\kappa_1+\kappa_2}{\kappa_1-\kappa_2}\right|.
\end{equation}
Equations (\ref{t6-36.6}) make it possible to express the
renormalized velocities in terms of the densities of soliton gases:
\begin{equation}\label{t7-17.20}
\begin{split}
  s_{1}=\frac{\kappa_1^2(1-\al_2f_{1})-\kappa_2^2\al_1f_{2}}{1-\al_1f_{2}-\al_2f_{1}},\\
    s_{2}=\frac{\kappa_2^2(1-\al_1f_{2})-\kappa_1^2\al_2f_{1}}{1-\al_1f_{2}-\al_2f_{1}}.
  \end{split}
\end{equation}
If, however, we express densities  $f_{1,2}$ from Eqs.~(\ref{t6-36.6}) in terms of velocities $s_{1,2}$,
\begin{equation}\label{t6-36.8}
  f_1=\frac{s_2-\kappa_2^2}{\al_2(s_2-s_1)},\quad f_2=\frac{\kappa_1^2-s_1}{\al_1(s_2-s_1)},
\end{equation}
and substitute these expressions into Eq.~(\ref{t6-35.5}), this system can be reduced to a remarkably simple diagonal
form:
\begin{equation}\label{t6-36.9}
  \frac{\prt s_1}{\prt t}+s_2\frac{\prt s_1}{\prt x}=0,\quad
  \frac{\prt s_2}{\prt t}+s_1\frac{\prt s_2}{\prt x}=0,
\end{equation}
where renormalized velocities $s_1,s_2$ are Riemann
invariants of this system of equations of the hydrodynamic type.

Formally, Eqs.~(\ref{t6-36.9}) resemble the gas dynamic
equations in the Riemann form, but physical properties of the soliton gas differ significantly from the
properties of a conventional gas. It should be noted
above all that although the soliton velocities are renormalized in the overlap region of soliton clouds, such a
change in velocity cannot be interpreted as the acceleration of solitons under the action of pressure: after
the clouds leave the overlap region, their velocities
restore their initial values. Further, the gases flow
freely through each other without experiencing any
dissipation processes. In this case, dynamic equations (\ref{t6-35.5}) 
have the form of conservation laws and can have
discontinuous solutions like in the theory of viscous
shock waves. Finally, we can assume in our case that
soliton gases have zero temperature so that a transition
through a discontinuity is not associated with an
increase of entropy; i.e., the Jouguet-Zempl\'{e}n
theorem (see, for example, \cite{LL6}) is inapplicable, and
the discontinuity can have any sign. This also means
that the existence of such formally discontinuous
flows is not associated with soliton pulse breaking followed 
by formation of a narrow transition layer
connecting both flows with different parameters like
in the theory of viscous shock waves. 

This situation is
clarified with the help of the interrelation between
Eqs.~(\ref{t6-36.9}) and the Chaplygin gas theory. Chaplygin \cite{chaplygin} 
noted that the equation of state of a gas,
\begin{equation}\label{t7-24.1}
  p=A-\frac{B}{\rho}
\end{equation}
($p$ is the gas pressure, $\rho$ is its density, and $A,B$ are constant parameters) 
can serve as a convenient approximation of small segments of the Poisson adiabat,
where the formulas of the theory are simplified significantly. (Chaplygin noticed also
the connection between the gas dynamic equations for this case and the theory of minimal surfaces.)
However, we approach the Chaplygin gas theory from a different point of view.

Already at the initial stage of development of the theory of shock waves, Stokes and Kelvin
discussed whether gas dynamic equations
\begin{equation}\label{t7-25.4}
  \rho_t+(\rho u)_x=0,\quad u_t+uu_x+\frac{c^2}{\rho}\rho_x=0,
\end{equation}
($c^2=dp/d\rho$, $c$ being the velocity of sound) permit solutions in the form of 
a traveling stationary wave.
With such a formulation of the problem, both density
$\rho$ and the flow velocity $u$ in the stationary solution are
obviously functions of only the coordinate $\xi=x-Vt$, these quantities are connected by a 
one-to-one dependence. Consequently, such a solution must be a simple
wave expressed by the Poisson relation (see \cite{LL6})
\begin{equation}\label{99-2}
    \rho=\rho_0[x-(c+u)t],
\end{equation}
$\rho_0(x)$ being the density profile in such a wave propagating with velocity $c+u$.
In this case, the relation between $\rho$ and $u$ is expressed by the condition of 
constancy of the Riemann invariant: $u-\int cd\rho/\rho=r_-=\mathrm{const}$ (see \cite{LL6}).
A wave with a stationary profile can exist only if
condition $c+u=\mathrm{const}$ is satisfied. Substituting into
this expression relations
$$
c=\sqrt{\frac{dp}{d\rho}},\quad u=\int_0^{\rho}\sqrt{\frac{dp}{d\rho}}\frac{d\rho}{\rho}
+\mathrm{const},
$$
and differentiating with respect to $\rho$, we obtain the following equation for 
gas equation of state $p=p(\rho)$
which permits a stationary wave:
$$
\frac{d^2p}{d\rho^2}+\frac2{\rho}\frac{dp}{d\rho}=0.
$$
This equation can be solved easily and gives exactly above dependence (\ref{t7-24.1}).
Setting for simplicity $B=1$, we obtain $c^2=dp/d\rho=1/\rho^2$, $c=1/\rho$ and the 
Riemann invariants (see \cite{LL6}) turn out to be
\begin{equation}\label{t7-24.2}
\begin{split}
  &s_1=u+\int\frac{cd\rho}{\rho}=u-\frac1{\rho},\\
  &s_2=u-\int\frac{cd\rho}{\rho}=u+\frac1{\rho},
  \end{split}
\end{equation}
where $u$ and $\rho$ obey the Euler equations for the Chaplygin gas:
\begin{equation}\label{t7-25.4}
  \rho_t+(\rho u)_x=0,\qquad u_t+uu_x+\frac{\rho_x}{\rho^3}=0,
\end{equation}
and can be expressed in terms of the Riemann invariants:
\begin{equation}\label{t7-24.3}
  \rho=\frac2{s_2-s_1},\qquad u=\frac12(s_2+s_1).
\end{equation}
If we write Eqs.~(\ref{t7-25.4}) in terms of Riemann invariants (\ref{t7-24.2}),
simple calculations lead to the dynamic equations for a Chaplygin gas in the diagonal Riemann
form coinciding with Eqs.~(\ref{t6-36.9}) which
describe the dynamics of two interacting soliton gases.

If we now seek the solution to Eq.~(\ref{t7-25.4}) in the form
of a wave traveling with constant velocity $V$  ($\rho=\rho(\xi),u=u(\xi)$, $\xi=x-Vt$),
we can easily see that these equations are satisfied for any function $\rho(\xi)$, if the 
Chaplygin gas flow velocity can be expressed in terms of $\rho(\xi)$ as
\begin{equation}\label{t7-25.7}
  u(\xi)=V+\frac1{\rho(\xi)}.
\end{equation}
In this case, the Riemann invariants are given by
\begin{equation}\label{t7-25.9}
  s_1=V,\qquad s_2=V+\frac2{\rho(\xi)}.
\end{equation}
These expressions obviously give a solution to
Eq.~(\ref{t6-36.9}) in the form of a simple wave: if we set, for
example, $s_1=V=\mathrm{const}$, the second equation is transformed into $s_{2,t}+Vs_{2,x}=0$ 
with general solution $s_2=F(x-Vt)$, which coincides with (\ref{t7-25.9}), if we write function $F(\xi)$
in form $F(\xi)=V+2/\rho(\xi)$. In particular, the function $F(\xi)$ can include a discontinuity 
propagating without change of form in the overlap region of the two gases.

The relation between the equations for a Chaplygin
gas and the kinetic equation describing the dynamics
of two interacting soliton clouds makes it possible to
construct instructive examples of solutions to the
kinetic equation.

\section{Collision of two soliton gases}

\begin{figure}[t]
\begin{center}
\includegraphics[width=7cm]{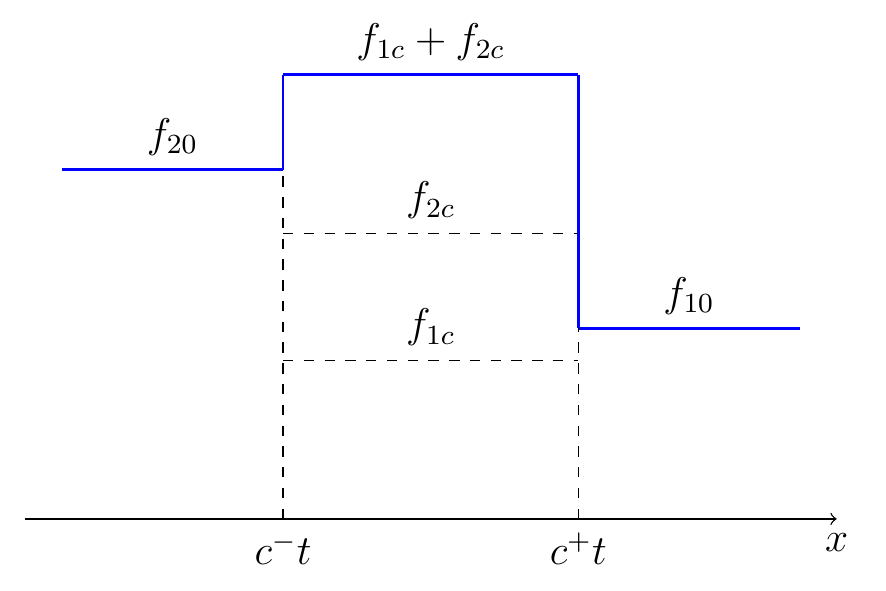}
\caption{
Coordinate dependence of the densities of soliton
gases interacting after a collision of soliton clouds with
constant initial densities and velocities.
 }
\label{fig11-15}
\end{center}
\end{figure}

Equations (\ref{t6-36.9}) have an obvious degenerate solution, in which $s_1$ and $s_2$
are constants depending neither on $x$, nor $t$. Despite the triviality of this solution, 
it has a clear physical meaning: it can serve as a
``plateau'' connecting two aforementioned solutions in
the form of simple waves. In particular, during the collision of mutually penetrating 
gases with constant densities $f_1,f_2$ and, hence, with Riemann invariants $s_1,s_2$, 
a region of a two-stream flow is formed, in
which the velocities of solitons of one species are
renormalized because of their interaction with solitons
of the other species. Thus, the problem of collision of
two gases is reduced to the determination of densities
and velocities of solitons in the region of their ``mixing'', 
as well as the velocities of the edges of this region.
Although this problem has already been analyzed in \cite{elkamch-05,cde-16}, 
we will briefly consider this problem because
the relevant results will be used in further analysis. We
assume that at the initial instant, there is a discontinuity in the density 
distributions for solitons of two species:
\begin{equation}\label{t6-37.16}
  \left\{
  \begin{array}{lll}
  f_2(x,0)=f_{20},\qquad & f_1(x,0)=0,\qquad &x<0,\\
  f_2(x,0)=0,\qquad & f_1(x,0)=f_{10},\qquad &x>0,
  \end{array}
  \right.
\end{equation}
here, we assume that $\kappa_2>\kappa_1$, so that the faster soliton
cloud with density $f_{20}$ overtakes at instant $t=0$  the
slower cloud with density $f_{10}$ at point $x=0$ and their
mutual penetration begins. As usual, in the case of initial discontinuity 
the solution must depend only on
self-similar variable $\zeta=x/t$. Therefore, the initial 
discontinuity leads to the formation of a plateau between
two simple waves, each of which is a discontinuity with
a constant value of one of the Riemann invariants (the
existence of such solutions is determined by the aforementioned properties 
of the Chaplygin gas). In the
whole, the solution is a sequence of three flows with
constant densities, which are separated by discontinuities:
\begin{equation}\label{t6-38.17}
  f(x,t)=\left\{
  \begin{array}{ll}
  f_{20},\qquad & x<c^-t,\\
  f_{1c}+f_{2c},\qquad & c^-t<x<c^+t,\\
  f_{10},\qquad & x>c^+t.
  \end{array}
  \right.
\end{equation}
It can clearly be seen from Fig.~\ref{fig11-15} that the leading
edge of the plateau moves with renormalized velocity
$s_{2c}$ of the fast gas, while the rear edge of the plateau
moves with renormalized velocity $s_{1c}$
of the slow gas because the densities at these edges vanish:
\begin{equation}\label{t6-38.21}
  c^-=s_{1c},\qquad c^+=s_{2c}.
\end{equation}
In the two-flow region $c^-t<x<c^+t$ of the plateau,
the soliton velocities are renormalized by their interaction; 
in accordance with relations (\ref{t7-17.20}) we have
\begin{equation}\label{t6-38.18}
\begin{split}
  &s_{1c}=\frac{\kappa_1^2(1-\al_2f_{1c})-\kappa_2^2\al_1f_{2c}}{1-\al_1f_{2c}-\al_2f_{1c}},
  \\
  &s_{2c}=\frac{\kappa_2^2(1-\al_1f_{2c})-\kappa_1^2\al_2f_{1c}}{1-\al_1f_{2c}-\al_2f_{1c}}.
  \end{split}
\end{equation}
The slow gas flows through the right discontinuity
into the plateau region; equating the expressions for its
flux on different sides of the discontinuity in the reference frame in which 
it is at rest, we obtain $f_{1c}(s_{1c}-c^+)=f_{10}(\kappa_1^2-c^+)$.
Fast gas flows through the left discontinuity into the plateau region, 
and analogous calculation gives
$f_{20}(\kappa_2^2-c^-)=f_{2c}(s_{2c}-c^-)$. With account for relations 
(\ref{t6-38.21}) and (\ref{t6-38.18})
the resulting equalities lead to equations
\begin{equation}
\begin{split}
&f_{1c}=f_{10}\frac{s_{2c}-\kappa_1^2}{s_{2c}-s_{1c}}=f_{10}(1-\al_1f_{2c}), \\
&f_{2c}=f_{20}\frac{\kappa_2^2-s_{1c}}{s_{2c}-s_{1c}}=f_{20}(1-\al_2f_{1c}),
\end{split}\nonumber
\end{equation}
which gives expressions
\begin{equation}\label{t6-38.22}
\begin{split}
  &f_{1c}=\frac{f_{10}(1-\al_1f_{20})}{1-\al_1\al_2f_{10}f_{20}},\\
  &f_{2c}=\frac{f_{20}(1-\al_2f_{10})}{1-\al_1\al_2f_{10}f_{20}},
  \end{split}
\end{equation}
for the densities of the soliton gases in the two-flow
region of the plateau. Clearly, these expressions have
sense only when the following inequality holds:
\begin{equation}\label{t6-38.23}
  \al_1f_{1c}+\al_2f_{2c}<1,
\end{equation}
when renormalized velocities (\ref{t6-38.18}) are positive, i.e., the
densities of soliton gases cannot be too high. Actually,
more stringent limitations on the soliton density are
imposed by the condition that variation $\overline{u^2}-\overline{u}^2$ of the
fluctuating wave field in a soliton gas must be positive
(see \cite{el-16}). The relations derived above are in good
agreement with the results of numerical solution of the
KdV equation with the initial data in the form of a
large number of solitons of two different species,
which are located on different sides of the coordinate
origin (see \cite{cde-16}).

\section{Flow of two soliton gases in the form of a simple wave}

\begin{figure}[t]
\begin{center}
\includegraphics[width=7cm]{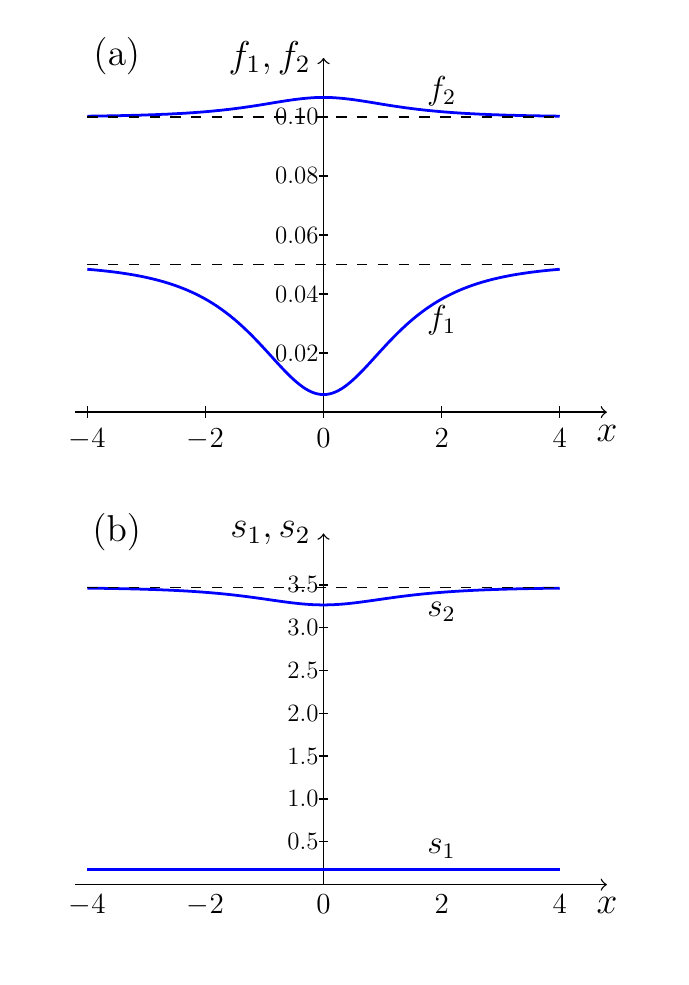}
\caption{
(a) Distributions of densities of soliton gases and (b)
of their renormalized velocities as functions of the coordinate for 
the wave profile defined by formula (\ref{t7-26.24}). Soliton
gas parameters: $\kappa_1=1$, $\kappa_2=1.8$, $f_{10}=0.05$, $f_{20}=0.1$. 
The wave velocity coinciding with the flow velocity of slow
solitons is $V=s_1=0.174$, while the flow velocity for the fast
soliton gas at infinity is $s_{20}=3.469$, which is higher than
nonrenormalized velocity $\kappa_2^2=3.24$.
 }
\label{fig11-16}
\end{center}
\end{figure}

Expressions (\ref{t7-25.9}) give a more general form of the
solution for the flow of two soliton gases in the form of
a simple wave. Let us suppose that the densities of the
soliton gases at infinity tend to constant values:
\begin{equation}\label{t7-26.15}
  f_1\to f_{10},\qquad f_2\to f_{20}\qquad \text{as}\qquad x\to\pm\infty.
\end{equation}
Then the constant value of Riemann invariant $s_1$ is
equal to the wave velocity of the soliton gas (see (\ref{t6-38.18}))
\begin{equation}\label{t7-26.22}
  V=s_{1}=\frac{\kappa_1^2(1-\al_2f_{10})-\kappa_2^2\al_1f_{20}}{1-\al_1f_{20}-\al_2f_{10}},
\end{equation}
which coincides with the renormalized velocity of the
slow gas, while formulas (\ref{t6-36.8}) give expressions for the
densities of soliton gases:
\begin{equation}\label{t7-25.10}
\begin{split}
  &f_1(\xi)=\frac1{\al_2}\left[1+\frac12(V-\kappa_2^2)\rho(\xi)\right],\\
  &f_2(\xi)=\frac1{2\al_1}(\kappa_1^2-V)\rho(\xi).
  \end{split}
\end{equation}
It can easily be seen that these densities are connected by relation
\begin{equation}\label{t7-25.11}
  1-\al_1f_{2}-\al_2f_{1}=\frac12(\kappa_2^2-\kappa_1^2)\rho,
\end{equation}
which implies, in particular, that
\begin{equation}\label{t7-26.21}
  \rho_0=\frac{2(1-\al_1f_{20}-\al_2f_{10})}{\kappa_2^2-\kappa_1^2}.
\end{equation}
Density distributions (\ref{t7-25.10}) contain arbitrary function $\rho(\xi)$ and 
the distribution of the renormalized
velocity of the fast soliton gas can also be expressed in
terms of this function in accordance with the second
expression in (\ref{t7-25.9}). At infinity, $s_2$ tends to the value (see
expressions (\ref{t6-38.18}))
\begin{equation}\label{t7-27.25}
  s_{20}=\frac{\kappa_2^2(1-\al_1f_{20})-\kappa_1^2\al_2f_{10}}{1-\al_1f_{20}-\al_2f_{10}}.
\end{equation}
The resulting solution has physical sense for such
values of parameters $\kappa_1,\kappa_2,f_{10},f_{20}$,
that the distribution of densities $f_1,f_2$ of the soliton gases, as well
as their renormalized velocities $s_1,s_2$ are positive everywhere.

Figure \ref{fig11-16} illustrates the flow for two soliton gases for
function $\rho(\xi)$  given by
\begin{equation}\label{t7-26.24}
  \rho(\xi)=\rho_0+\frac{a}{\cosh\xi}.
\end{equation}
In this case, the profile of density $f_1$ of the slow gas
moves together with this component with velocity
$V=s_1$. The fast gas flows through the slow gas with local
velocity $s_2$, which decreases in the region of the ``well''
in distribution $f_1$, this leads to an increase in density $f_2$ in this region.

\section{General solution for the flow of two soliton gases}

The hodograph method, which is well known in
gas dynamics (see, for example, \cite{LL6}) makes it possible to easily 
obtain the general solution to Eqs.~(\ref{t6-36.9}), in
the case when both velocities $s_1$ and $s_2$ vary in space
and time. Performing standard hodograph transformation $t=t(s_1,s_2)$, $x=x(s_1,s_2)$,
we reduce these equations to the form
\begin{equation}\label{t6-44.1}
  \frac{\prt x}{\prt s_1}-s_1\frac{\prt t}{\prt s_1}=0,\qquad
  \frac{\prt x}{\prt s_2}-s_2\frac{\prt t}{\prt s_2}=0.
\end{equation}
Eliminating $x$, we obtain
$$
\frac{\prt^2t}{\prt s_1\prt s_2}=\frac1{s_1}\frac{\prt^2x}{\prt s_1\prt s_2}=
\frac1{s_1}\frac{\prt}{\prt s_1}\left({s_2}\frac{\prt t}{\prt s_2}\right)=
\frac{s_2}{s_1}\frac{\prt^2t}{\prt s_1\prt s_2}.
$$
Since $s_1\neq s_2$, we arrive at equation
\begin{equation}\label{t6-44.2}
  \frac{\prt^2t}{\prt s_1\prt s_2}=0,
\end{equation}
the general solution to which can be written in terms of
two arbitrary functions (it is convenient to denote them as
$F^{\prime\prime}(s_1)$ and $G^{\prime\prime}(s_2)$), in form
\begin{equation}\label{t6-44.3}
  t(s_1,s_2)=F^{\prime\prime}(s_1)+G^{\prime\prime}(s_2).
\end{equation}
Equations (\ref{t6-44.1}) readily give
\begin{equation}\label{t6-44.4}
  x(s_1,s_2)=s_1F^{\prime\prime}(s_1)+s_2G^{\prime\prime}(s_2)- F'(s_1)- G'(s_2).
\end{equation}
This solution can be transform to
\begin{equation}\label{t6-44.6}
  \begin{split}
  & x-s_1t=(s_2-s_1)G^{\prime\prime}(s_2)- F'(s_1)- G'(s_2),\\
  & x-s_2t=-(s_2-s_1)F^{\prime\prime}(s_1)- F'(s_1)- G'(s_2).
  \end{split}
\end{equation}
In this form, the solution has been obtained in \cite{pavlov-87b} by
a different method.

For completeness of analysis, we consider other
representations of the general solution, which may
turn out to be more convenient for solving specific
problems. If we turn to the hodograph method in the
Tsarev form \cite{tsarev-90}, the general solution can be written in the form
\begin{equation}\label{t6-36.10}
  x-s_2t=\frac{\prt W}{\prt s_1},\qquad x-s_1t=\frac{\prt W}{\prt s_2},
\end{equation}
where function $W=W(s_1,s_2)$ must satisfy the Euler-Poisson equation:
\begin{equation}\label{t6-36.11}
  \frac{\prt^2W}{\prt s_1\prt s_2}+
    \frac{1}{s_1-s_2}\left(\frac{\prt W}{\prt s_1}-\frac{\prt W}{\prt s_2}\right)=0.
\end{equation}
It can easily be verified that the solutions to this
equation can be expressed in terms of arbitrary functions $F(s_1)$ and $G(s_2)$ as follows:
\begin{equation}\label{t6-36.12}
  W(s_1,s_2)=(s_1-s_2)[F'(s_1)-G'(s_2)]-2[F(s_1)+G(s_2)].
\end{equation}
Substituting this expression into (\ref{t6-36.10}), we obtain the
solution in form (\ref{t6-44.6}).
Functions $F(s_1)$ and $G(s_2)$ should be determined from the initial conditions to the
problem.

As was shown by Riemann (see, for example, \cite{sommer-50}), instead of using this 
general solution expressed in
terms of arbitrary functions, it is often convenient to
employ the method in which the solution to the problem is expressed directly 
in terns of initial conditions
with the help of the so-called Riemann function. In
our case, the Riemann function can be obtained from
the general expression for the polytropic gas dynamics
with adiabatic exponent $\ga=-1$. The hypergeometric
function appearing in this expression reduces
to $F(-1,2;1;z)=1-2z$ and the Riemann function
for Eqs.~(\ref{t6-36.9}) takes an especially simple form:
\begin{equation}\label{t6-37.15}
  R(r_1,r_2;s_1,s_2)=\frac{(r_1+r_2)(s_1+s_2)-2(r_1r_2-s_1s_2)}{(r_1-r_2)^2},
\end{equation}
where $(r_1,r_2)$  are the current coordinates on the hodograph plane and $(s_1,s_2)$
are the coordinates of point $P$, at which the value of function $W=W(P)=W(s_1,s_2)$ is
sought. We will not write here rather cumbersome
general expressions that do not add much for understanding the dynamics of interacting soliton gases;
instead of this, we consider the solution to one of typical gas dynamic problems, viz., 
the problem of expansion of a gas layer. The solution can easily be obtained
in elementary form, and its difference from the solutions to the same problem for other physical systems
demonstrates well the substantial difference between
the conventional gas dynamics and the dynamics of soliton gases.

\section{Expansion of a layer from a mixture of two soliton gases }

One of typical problems in which a domain of the
general solution appears is the problem of expansion
of a gas layer. This problem was considered in \cite{nk-76} for
a classical monatomic gas; in \cite{ik-19b} for the Bose-Einstein condensate; 
in \cite{ik-20} for a two-temperature
plasma, and in \cite{landau-53,khal-54,kamch-19}
for an ultrarelativistic gas. In
all cases, rarefaction waves propagated from the edges to
the bulk of the layer, and after their collision, the general solution 
domain bordering the rarefaction waves at
the edges was formed. However, in the case of a soliton
gas, there is no solution in the form of rarefaction
waves, and instead of this, discontinuous solutions
separated by a plateau with constant values of the Riemann invariants appears. 
The solution constructed in
this way makes it possible to draw certain general conclusions about 
the evolution of overlapping clouds of soliton gases.

If we assume that a single soliton gas with parameter
$\kappa_1$ at the initial moment of time has a density distribution in
the form of a plateau with value $f_{10}$ for
$-l/2\leq x\leq l/2$, its motion at subsequent instants is obvious: 
this distribution is transferred along the $x$ axis with velocity $\kappa_1^2$ 
without change of its shape. If, however, there is a
mixture of soliton gases with parameters $\kappa_1$ and $\kappa_2$, $(\kappa_1<\kappa_2)$
and densities $f_{10}$ and $f_{20}$ at the initial instant in
domain $-l/2\leq x\leq l/2$, then the evolution of these gases is
more complex: in the course of separation of these
gases into two clouds moving with velocities $\kappa_1^2$ and $\kappa_2^2$, 
a two-flow plateau appears. Our aim is the description
of the process of spatial separation of soliton clouds
and the determination of their parameters after the separation.

\begin{figure}[t]
\begin{center}
\includegraphics[width=7cm]{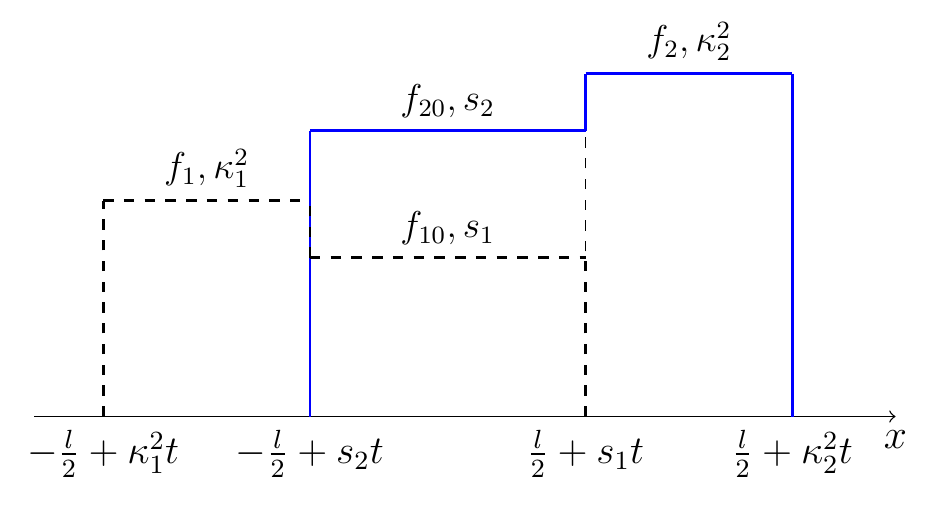}
\caption{
Coordinate dependencies of the densities of two
soliton gas clouds during the evolution of the initial layer of
two gases before the instant of their separation. Bold
dashed line shows the density of the slow gas, while solid
line corresponds to the density of the fast gas.
 }
\label{fig11-17}
\end{center}
\end{figure}

\begin{figure}[t]
\begin{center}
\includegraphics[width=7cm]{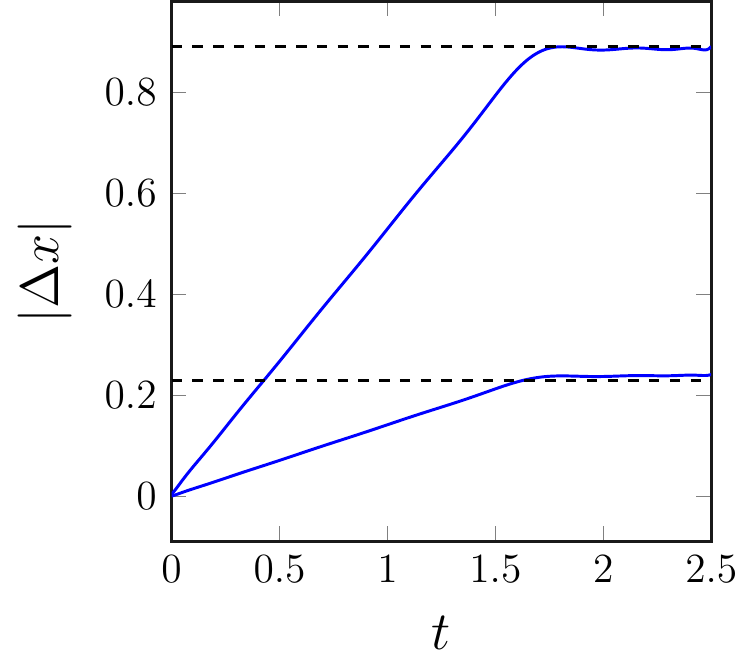}
\caption{
Shifts of the centers of mass of clouds from their
positions in the absence of interaction, which are obtained
by numerical solution of equations for the Chaplygin gas
for the initial values of parameters
$f_{10}=0.05$, $f_{20}=0.1$, $\kappa_1=1.3$, $\kappa_2=2.5$, $l=10$. At instant $T=1.6$
these parameters assume the theoretical values given by
formulas (\ref{t7-28.5}).
 }
\label{fig11-18}
\end{center}
\end{figure}

On the plateau formed during the separation, the
gases obviously move with renormalized velocities $s_1, s_2$, preserving their 
initial densities $f_{10},f_{20}$. In front of this layer, there appears a fast gas layer with
density $f_2$, while behind this layer, a slow gas layer with density $f_1$ is formed. The left edge
of the overlap region moves with velocity $s_2$, while its
right edge moves with velocity $s_1$. Consequently, the overlap region exists during time
\begin{equation}\label{t7-27.1}
  T=\frac{l}{s_2-s_1}
\end{equation}
and the gases are separated at $t>T$. It can easily be
seen from Fig.~\ref{fig11-17}, that the lengths of the clouds of slow
and fast gases at the instant of separation are given by
\begin{equation}\label{t7-27.2}
  l_1=l\cdot\frac{s_2-\kappa_1^2}{s_2-s_1}, \qquad l_2=l\cdot\frac{\kappa_2^2-s_1}{s_2-s_1},
\end{equation}
respectively, both of them being smaller than the initial
length $l$, because $s_1<\kappa_1^2$ and $s_2>\kappa_2^2$.
From the conservation of the number of solitons in each component, we find amplitudes
\begin{equation}\label{t7-28.3}
  f_1=f_{10}\cdot\frac{s_2-s_1}{s_2-\kappa_1^2},\qquad f_2=f_{20}\cdot\frac{s_2-s_1}{\kappa_2^2-s_2}.
\end{equation}
At the instant of separation of the clouds, the centers of mass of the slow soliton cloud 
and the fast soliton cloud are, respectively, at points
\begin{equation}\label{t7-28.4}
  x_{1}(T)=\frac{l}2\cdot\frac{s_1+\kappa_1^2}{s_2-s_1},\qquad
  x_{2}(T)=\frac{l}2\cdot\frac{s_2+\kappa_2^2}{s_2-s_1}.
\end{equation}
If solitons did not interact with one another, these
clouds would move with nonrenormalized velocities
and would be at points
$x_{10}(T)=\kappa_1^2l/(s_2-s_1)$, $x_{20}(T)=\kappa_2^2l/(s_2-s_1)$ at this instant; 
i.e., because of the interaction, the clouds are shifted through distances
\begin{equation}\label{t7-28.5}
\begin{split}
  &\Delta x_1=x_1(T)-x_{10}(T)=\frac{l}{2}\cdot\frac{s_1-\kappa_1^2}{s_2-s_1}<0,\\
  &\Delta x_2=x_2(T)-x_{20}(T)=\frac{l}{2}\cdot\frac{s_2-\kappa_2^2}{s_2-s_1}>0.
  \end{split}
\end{equation}
Therefore, because of their interaction, the clouds
are narrowed, the soliton densities in both clouds
increase, and the slow cloud is shifted backwards,
while the fast cloud is shifted in the forward direction.
A transition to these values of the shifts are illustrated
in Fig.~\ref{fig11-18}, which is obtained from the numerical solution for 
the Chaplygin gas. At the initial instant, the
centers of mass of both gases were at the origin of coordinates, 
and in the course of separation of the clouds,
the center of mass of the fast gas moved more rapidly
than it would do with its nonrenormalized velocity,
while the center of mass of the slow gas would lag
behind the analogous motion with its nonrenormalized velocity. 
After the instant of separation, the centers of mass of both gases continue their motion with
nonrenormalized velocities, acquiring additional
shifts due to the interaction with solitons of other species. Good agreement 
between the values of shifts (\ref{t7-28.5}) predicted by the theory and the result of numerical
solution means that our idea of the formation of simple waves with discontinuities from 
the initial discontinuities corresponds to the actual dynamics of interacting soliton clouds 
in accordance with the kinetic equation.

\section{Conclusion}

In this study, it is shown that a fundamental property of the dynamics of two interacting soliton gases is
the existence of simple waves propagating without a
change in form. In particular, such simple waves can
have discontinuities; therefore, in problems of the
``dam breaking'', instead of rarefaction waves in the
dynamic of conventional gases, simple waves appear
with discontinuities, which are connected (instead of
the general solution in which both Riemann invariants
change) by a plateau region with constant Riemann
invariants. This property of the dynamics of soliton
gases differ drastically from the properties of gas
dynamics of conventional gases. At the same time,
based on the aforementioned pattern, simple analytic
solutions to typical problems, which are confirmed by
exact numerical solutions of the kinetic equation
reduced to the equivalent form of equations for the
dynamics of the Chaplygin gas, can be constructed.
The equations constructed in this way lead to an intuitively clear pattern of the behavior of soliton gas
clouds: as a result of the interaction, the cloud of fast
solitons is shifted in the forward direction, while the
cloud of slow solitons is shifted backwards, both
clouds becoming narrower and soliton densities
increase in both of them. The resulting expressions make it
possible to estimate these effects in current experimental investigations of soliton gases in waves on the
water surface, in nonlinear optics, and in the physics
of cold atoms (see, for example, \cite{redor-19,suret-20,bd-22}).

The authors are grateful to G.~A.~ El for fruitful discussions. This study was supported by the Russian Foundation for
Basic Research (project no.~20-01-00063).

\end{document}